\documentclass[a4paper,12pt]{article}

\usepackage{amsmath,amssymb,graphicx}
\usepackage{pst-all}
\usepackage{cite}

\numberwithin{equation}{section}

\newcommand{\ii}{\mathrm{i}}

\newcommand{\dd}{\mathrm{d}}

\newcommand{\Pb}{\mathcal{P}}

\newcommand{\e}{\mathrm{e}}

\newcommand{\tr}{\mathop{\mathrm{tr}}\nolimits}
\renewcommand{\Tr}{\mathop{\mathrm{Tr}}\nolimits}

\newcommand{\ft}[2]{{\textstyle\frac{#1}{#2}}}
\newcommand{\qp}[1]{[\!\![ #1 ]\!\! ]}

\begin{document}

\title{Random walks in $\mathcal{N}=4$ Super Yang--Mills}%
\author{Corneliu Sochichiu\thanks{Address after 01.01.2007: Laboratori Nazionali di Frascati, via E.Fermi 40, I-00044 Frascati (RM), Italy} \\
 {\it Max-Planck-Institut (Werner-Heisenberg-Institut)}\\
 {\it F\"{o}hringer Ring, 6, D-80805 Munich, GERMANY}\\
 and\\
 {\it Institutul de Fizic\u{a} Aplicat\u{a}}\\
 {\it str.Academiei, 5,
 MD-2028 Chi\c{s}in\u{a}u, MOLDOVA}\\
e-mail: \texttt{sochichi@lnf.infn.it}
}%


\maketitle
\begin{abstract}
 Using an effective description of the thermal partition function for SU(2) sector of $\mathcal{N}=4$ super Yang--Mills theory in terms of interacting random walks we compute the partition function in planar limit as well as give the leading non-planar contribution. The result agrees with existent approaches in what concerns the zero coupling and one-loop Hagedorn temperature computation.
\end{abstract}

\section{Introduction}

Gauge / string correspondence has a long history. In \cite{'tHooft:1973jz} 't Hooft proposed to reformulate the gauge theory perturbative expansion in terms of a topological expansion with the parameter of expansion given by $1/N$, the inverse rank of the gauge group. Since that started the hunt of string theory in the large $N$ gauge theory. The most success was reached in the framework of the maximally supersymmetric ($\mathcal{N}=4$) Yang--Mills (SYM) theory, where an equivalence between the gauge theory at infinite $N$ and the theory of strings on the AdS$_5 \times S^5$ was explicitly claimed. This conjecture known under the name AdS/CFT correspondence \cite{Maldacena:1997re}. After the conjecture has been made there were many supporting tests of it, but no rigorous proof so far. (See \cite{Aharony:1999ti} for an extensive classical review of the subject and \cite{Plefka:2005bk} for a more recent progress.)

The above conjecture stimulated, among others, an intensive study of anomalous dimensions and mixing of SYM composite operators (see \cite{Beisert:2004ry} for a detailed review and references).
In particular, for some sectors of the SYM the quantum dilatation operator was constructed as a perturbative expansion to up to three loops in terms of some differential operators \cite{Beisert:2003jj,Beisert:2003tq}. In this work we consider the one-loop level of SU(2) sector.

The system consisting of composite operators and dilation operator can be mapped to a spin system which in addition to the spin chain dynamics possesses an interaction responsible for chain splitting/joining \cite{Bellucci:2004ru,Bellucci:2004qx,Bellucci:2004dv}. The same system can be described in terms of a matrix mechanical model \cite{Bellucci:2004fh,Agarwal:2004cb}.

Another ``hot topic'' is the extension of the AdS/CFT correspondence to a finite temperature \cite{Witten:1998zw}.
This extension allows one to identify the confinement / deconfinement phase transition in the gauge theory with the Hagedorn transition on the string side. In this context it is interesting to consider a nonzero temperature on the space of composite SYM operators as well \cite{Sochichiu:2006uz}. Since the composite operators in $\mathcal{N}=4$ SYM are identified through AdS/CFT correspondence with the quantum states in string theory this corresponds to thermal strings on AdS$_5 \times S^5$ \cite{Witten:1998zw}. On the other hand, since the dilatations in a conformal theory can be identified with time evolution this should be equivalent to SYM itself at finite temperature.
Indeed, the analysis shows that this is qualitatively compatible with the results of thermal Yang--Mills theory in four dimensions \cite{Aharony:2005bq,Aharony:2003sx}.

In \cite{Sochichiu:2006uz} the author proposed to treat $N$ as a thermodynamically large rather than an infinite number. The advantage of such an approach is that one can get the reliable information about the system in both phases of the theory while $N$ is serving as both coupling and cut off parameter. In particular, it was found that below the Hagedorn critical point the system at large $N$ is independent of $N$, while beyond this point it scales as $N^2$. (Another phase with the scaling $N$ to the power one was suggested to exist for large $\beta g_{\rm YM}^2$, where $\beta$ is the inverse temperature.)

The aim of the present note is to refine and extend the analysis of \cite{Sochichiu:2006uz}, in special in what concerns the phase below the Hagedorn transition. As in
\cite{Sochichiu:2006uz} the thermal partition function is reduced to the integral over the eigenvalues of (compactified) time component of the matrix model gauge field.

Here we find it convenient to substitute the integration over the angular variables by averaging over \emph{random walks}. The random walks are defined in such a way that at each step the direction is chosen by the respective angle. In these terms the model looks like  a system of interacting random walks which in the limit $N\to\infty$ decouple from each other. (Let us note that the idea to describe a spin system, in particular Heisenberg spin chain in terms of random walks is not new, see e.g. \cite{Conlon:1991}.)

For sufficiently large $N$ the probability distribution for the walks is asymptotically Gaussian. This allows one to compute the partition function explicitly as a perturbation in 't Hooft coupling $\lambda$ and $1/N$. In particular we compute the zero coupling and one-loop correction to the Hagedorn value of temperature and chemical potential. The result which seems to be obtained at almost no cost matches perfectly the known results of P\'olya enumeration theorem \cite{Spradlin:2004pp}.

The plan of the paper is as follows. In the next section we briefly describe the thermal partition function for the matrix model and reduce it to an integral over $N$ angular variables. In the third section we reformulate the model in terms of random walks. In section four we compute the leading gaussian contribution to the partition function and find the criticality condition for the Hagedorn transition. Finally, we discuss the results.

\section{Partition function}

Consider the SU(2) sector of composite operators of $\mathcal{N}=4$ SYM. The effect of dilatations on these composite operators is given by the quantum dynamics of the matrix model described by the following action \cite{Bellucci:2004fh},
\begin{multline}\label{action}
  S(\Psi,\bar{\Psi})=\\
  \int\dd\tau\,
  \tr\left\{
         \frac{\ii}{2}(\bar{\Psi}_a\nabla_0\Psi_a-\nabla_0\bar{\Psi}_a\Psi_a)
         -\frac{ g_{\rm YM}^2}{16\pi^2}
         [\bar{\Psi}_a,\bar{\Psi}_b][\Psi^a,\Psi^b]
  \right\},
\end{multline}

Let us consider the grand canonical partition function,\footnote{While ``$\tr$'' denotes the trace over $N\times N$ matrices, ``$\Tr$'' denotes the one over the Hilbert space.}
\begin{equation}
  Z(\mu,\vec{x})=\Tr \e^{-\mu L-\vec{x}\cdot \vec{S}-\beta V},
\end{equation}
where $\mu$, $\vec{x}$ are respectively the chemical
potentials for conserved charges: the total number of excited modes alias total chain length operator,
\begin{equation}
  L=\tr\bar{\Psi}_a\Psi_a, \qquad a=1,2,
\end{equation}
total spin operator,
\begin{equation}
  \vec{S}=\ft12\tr\bar{\Psi}_a\vec{\sigma}_{ab}\Psi_b,
\end{equation}
and $\beta$ is inverse temperature. The non-quadratic interaction potential is known as a perturbation theory expansion in the Yang--Mills coupling $g_{\rm YM}$: $V=\sum_k g_{\rm YM}^{2k}V_{(k)}$. The first term of the expansion is
\begin{equation}
  V_{(2)}=\frac{g_{\rm YM}^2}{16\pi^2}\tr[\bar{\Psi}_a,\bar{\Psi}_b][\Psi_a,\Psi_b].
\end{equation}
Both $L$ and $\vec{S}$ are conserved quantities, i.e.
\begin{equation}
  \qp{L,V_{int}}=\qp{\vec{S},V}=\qp{L,\vec{S}}=0,
\end{equation}
where $\qp{\cdot,\cdot}$ denotes the quantum commutator to be
distinguished from the (classical) matrix one:  $[\cdot,\cdot]$, which denotes only permutations
in matrix indices but not of the operators. Next, for any given $\vec{x}$ we can choose indices
$a$ and $b$ to label the components along eigenvectors of $x_{ab}=\vec{x}\cdot\vec{\sigma}_{ab}$.
Then, the partition function can be rewritten in the following terms,
\begin{equation}\label{eq:z}
  Z(\mu,\vec{x})=\Tr \e^{-\beta V-\sum_{a=\pm}\mu_a L_a},\qquad a=\pm,
\end{equation}
and
\begin{equation}
  \mu_{\pm}=\mu\pm\frac{x}{2}, \qquad x=\sqrt{\vec{x}^2},
\end{equation}
i.e. the partition function depends only on the absolute value of $\vec{x}$, as it could be expected.

Consider the perturbative expansion of \eqref{eq:z}. The formal expression for the expansion can be encoded as,
\begin{equation}\label{pert}
  Z(\mu,\vec{x})=\Tr T_{\beta}\e^{-\beta H_0}\exp\left(-\int_0^\beta V_{\tau}\dd\tau\right),
\end{equation}
where,
\begin{align}
  H_0&=\beta^{-1}(\mu_{+} L_{+}+\mu_{-}L_{-}),\\
  V_{\tau}&=\e^{\beta H_0}V_{(2)}\e^{-\beta H_0},
\end{align}
and $T_{\beta}$ is the ``thermal-ordering operator'',
\begin{equation}
  T_{\beta}V_{\tau}V_{\tau'}=
  \begin{cases}
    V_{\tau}V_{\tau'}, & \tau < \tau',\\
    V_{\tau'}V_{\tau}, & \tau > \tau'.
  \end{cases}
\end{equation}
In our case the perturbation $V_{(2)}$ commutes with $H_0$, therefore instead of $V_\tau$ we can simply use $V_{(2)}$ as well as drop the $T_{\beta}$-ordering from the trace.

The first terms in the perturbation theory expansion read,
\begin{equation}\label{eq}
  Z(\mu,\vec{x})=\Tr\e^{-\beta H_0}(1-\beta V+\dots)\equiv
  Z_{0}(\mu,\vec{x})(1-\beta \langle V\rangle_0\dots),
\end{equation}
where $Z_0$ is the partition function for the gauged matrix oscillator, as well as the mean $\langle\cdot\rangle_0$ denotes one computed with respect to the gauged oscillator,
\begin{equation}
  \langle V\rangle_0=\frac{\Tr \e^{-\beta H_0}V}{\Tr\e^{-\beta H_0}}.
\end{equation}

In \cite{Sochichiu:2006uz} $Z_0(\mu,x)$ and $\langle V\rangle_0$ were computed in terms of an integral over eigenvalues $\theta_n$, $n=1,\dots,N$ of the (compactified) gauge field,
\begin{multline}\label{z0alpha}
  Z_0(\mu,\vec{x})=
  \frac{2^{-\ft12N(N+1)}\e^{N^2\mu}}{[\sinh(\mu_{+}/2)\sinh(\mu_{-}/2)]^{N}}
  \int\prod_n\dd\theta_n\,\times\\\prod_{m>n}
  \frac{1-\cos\theta_{mn}}
  {(\cosh\mu_{+}-\cos\theta_{mn})(\cosh\mu_{-}-\cos\theta_{mn})}
  ,
\end{multline}
and, respectively,
\begin{multline}\label{V:h}
  \langle V\rangle_{0,\theta}=
  \frac{\beta g_{\rm YM}^2}{8\pi^2}
  \left(\frac{N}{(\e^{\mu_{+}}-1)(\e^{\mu_{-}}-1)}
  \right.\\
  \left.-\ft14\sum_{knm}\frac{\cos\theta_{mk}-
  \e^{-\mu_{+}}\cos\theta_{nk}-\e^{-\mu_{-}}\cos\theta_{mn}+\e^{-2\mu}}
  {(\cosh\mu_{+}-\cos\theta_{mn})(\cosh\mu_{-}-\cos\theta_{nk})}\right),
\end{multline}
where the last equation gives the mean value of the potential in
fixed $\theta$-background.\footnote{To get the ``true'' mean one
should integrate the eq. \eqref{V:h} over $\theta_n$'s with a
measure given by the integrand of \eqref{z0alpha}.} Equations \eqref{z0alpha} and \eqref{V:h} are obtained by explicit evaluation of the, respectively, gaussian integral in $\Psi$ and its perturbation by $V_{(2)}$.

The partition function \eqref{z0alpha} and the mean \eqref{V:h} can be represented in the following form \cite{Sochichiu:2006uz},
\begin{equation}\label{z0mu-exp}
  Z_0(\mu,\vec{x})=\int[\dd\theta]\exp
  \left[
   -\sum_{\omega=1}^{\infty}\frac{1}{\omega}(1-
  \e^{-\omega \mu_{+}}-\e^{-\omega \mu_{-}})\sum_{mn}\e^{\ii\omega\theta_{mn}}
  \right],
\end{equation}
where $[\dd\theta]\equiv\prod_n(\dd\theta_n/2\pi)$, and,
\begin{equation}\label{v-mean-anlt}
  \langle V\rangle_{0,\theta}=
  \frac{\beta \lambda}{8\pi^2}
  \sum_{\omega,\omega'\geq 1}
  \e^{-\mu_{+}\omega -\mu_{-}\omega'}
  \left(1-\ft{1}{2N}\sum_{mnk}\e^{\ii\omega\theta_{mn}
  +\ii\omega' \theta_{nk}}\right)
  ,
\end{equation}
where $\lambda$ is the 't Hooft coupling: $\lambda=g_{\rm YM}^2N$.

Equations \eqref{z0mu-exp} and \eqref{v-mean-anlt} will serve as a
starting point of our present study.

\section{Random walk variable}

Let us introduce the field $\varphi_\omega$ defined as,\footnote{In
\cite{Sochichiu:2006uz} we considered the real part of
$\varphi_\omega$ under assumption of symmetric distribution of
$\theta_n$, which we called $\rho_\omega$. Here we relax this
assumption about $\theta_n$ distributions, and consider a complex
$\varphi_\omega$ instead.}
\begin{equation}
  \varphi_\omega=\sum_n\e^{\ii \omega \theta_n}.
\end{equation}

One can view the angles $\theta_m$ as random variables and
respective integrals as the averaging over such variable. In this
case the function $\varphi_\omega$ has the meaning of a position
after $N$ steps of random walk with $\omega\theta_n$ giving the random
direction chosen at step $n$.

The random walks $\varphi_\omega$ have the following properties:
\begin{subequations}\label{propties}
\begin{itemize}
   \item Except $\varphi_0=N$, all $\varphi_\omega$ have zero expectation values
   \begin{equation}\label{mean}
      \overline{\varphi_\omega }
     \equiv\int[\dd\theta]\sum_n\e^{\ii\omega\theta_n}=0,\qquad \omega\neq 0,
   \end{equation}
   where the wide bar denotes the mean with respect to
   constant $\theta$-distribution. (Not to be confused with complex conjugation denoted by a simple bar.)
   \item For different $\omega$'s the random walks have the pair correlators vanishing.
   The only non-zero pair correlator is given by,
   \begin{equation}\label{2corr}
     \overline{ \bar{\varphi}_\omega \varphi_{\omega'}}=N\delta_{\omega\omega'},
   \end{equation}
   where $\bar{\varphi}_\omega$ is the complex conjugate of $\varphi_\omega$.
   \item In general the higher order correlators are given by the quadratic correlators plus sub-leading terms. In particular, the $\bar{\varphi}\varphi\varphi$-correlator is given by a subleading term only
   \begin{equation}\label{3corr}
     \overline { \bar{\varphi}_\omega\varphi_{\omega_1}\varphi_{\omega_2}}=
     N\delta_{\omega,\omega_1+\omega_2},
   \end{equation}
   while the $\bar{\varphi}\bar{\varphi}\varphi\varphi$-correlator contains both gaussian contribution as well as subleading non-gaussian terms,
   \begin{multline}\label{4phi}
     \overline{\bar{\varphi}_{\omega_1}     \bar{\varphi}_{\omega_2}\varphi_{\omega'_1}
     \varphi_{\omega'_2}}=\\
     N(N-1)[\delta_{\omega_1,\omega_1'}\delta_{\omega_2,\omega_2'}+
     \delta_{\omega_1,\omega_2'}\delta_{\omega_2,\omega_1'}]\\
     +N\delta_{\omega_1+\omega_2,\omega_1'+\omega_2'}.
   \end{multline}
   The order $N^2$ term in the second line appears because of quadratic correlators \eqref{2corr} while the subleading term $\sim N$ is in the last line is due to non-quadratic contribution.
   \item For independent and distinct $\theta_n$'s ($\theta_n\neq\theta_m$ for $n\neq m$),
   first $N$ random walks $\varphi_\omega$ are functionally
   independent\footnote{Taking into account the constraint $\sum_n\theta_n=0$ the
   number of independent random walks reduces to $N-1$.} and a configuration
   of $\theta_n$'s can be equivalently described by the set of independent random
   walks $\{\varphi_\omega, \omega=1,\dots, N\}$.
\end{itemize}
\end{subequations}

Next, the idea is to replace the $\theta$-integral by averaging with respect to \emph{random walks},
\begin{equation}
  \int[\dd\theta]F(\{\varphi_\omega(\theta)\})\to
  \int[\dd\bar{\varphi}_\omega\dd\varphi_\omega]\Pb(\{\varphi_\omega\})
  F(\{\varphi_\omega\}),
\end{equation}
where $\Pb(\{\varphi_\omega\})$ is the probability density for $\varphi_\omega$ distribution. For a large number of steps $N$ the probability distribution $\Pb(\{\varphi_\omega\})$ can be figured out from the properties \eqref{propties}. Asymptotically it is given by
   \begin{equation}\label{phi-dist}
     \Pb(\{\varphi_\omega\})=\frac{1}{(\pi N)^N}\,
     \e^{-\ft1N\sum_{\omega=1}^{\infty}\bar{\varphi}_\omega\varphi_\omega
     -\ft{1}{N^2}
     \sum_{\omega_1,\omega_2}(\bar{\varphi}_{\omega_1+\omega_2}
     \varphi_{\omega_1}\varphi_{\omega_2}+c.c.)-\dots},
   \end{equation}
where $c.c.$ denote the complex conjugate and dots stand for terms of higher order in $1/N$. The first term in the exponent of \eqref{phi-dist} is also one predicted by the Central Limit Theorem while the next one is the correction responsible e.g. for the nonzero $\bar{\varphi}\varphi\varphi$-correlator \eqref{3corr}.

The asymptotic expansion \eqref{phi-dist} to the probability density is valid as soon as random walks
$\varphi_\omega$ do not depart too mach from the origin. Also note that this expansion does not take care of non-analytic terms in $1/N$, which can be related to the ``non-perturbative''  contribution. The Gaussian distribution can be regarded as a sort of ``soft'' cut-off of $\varphi_{\omega}$ by the level
$|\varphi_{\omega}|\sim \sqrt{N}$. There is also a ``hard'' cut-off of $\varphi_\omega$ by,
  \[
  |\varphi_\omega|\leq N,
  \]
which follows from the definition of $\varphi_\omega$.
On the other hand, the distribution \eqref{phi-dist} gives although almost vanishing, but however
non-zero probability to find $\varphi_\omega$ beyond this
bound.

For the lower modes $\omega\ll N$, the variation of the measure is negligible with respect to the variation of the integrand of \eqref{z0alpha} rewritten in terms of $\varphi_{\omega}$, therefore for lower modes the probability distribution can be considered to be constant.

Now we are ready to rewrite the partition function in terms of random walks.
The one-loop partition function
\eqref{eq} with $Z_0$ and $\langle V\rangle_0$ given by
\eqref{z0mu-exp} and \eqref{v-mean-anlt} respectively can be written
as,
\begin{multline}\label{z-phi}
  Z_{(1)}(\mu,x)=\\
  \int[\dd\varphi\dd\bar{\varphi}]\exp
  \left\{
     -\sum_{\omega=1}^{\infty}\frac{1}{\omega}(1-
  \e^{-\omega \mu_{+}}-\e^{-\omega
  \mu_{-}})\varphi_{-\omega}\varphi_\omega
  \right.\\
  \left.
  +\frac{\beta\lambda}{8\pi^2}\sum_{\omega,\omega'}\e^{-\mu_{+}\omega-\mu_{-}\omega'}
  \left[1-\ft{1}{2N}(\varphi_{\omega}\varphi_{-\omega'}\varphi_{\omega'-\omega}
  +\varphi_{-\omega}\varphi_{\omega'}\varphi_{\omega-\omega'})
    \right]
  \right\}.
\end{multline}
The second line in \eqref{z-phi} corresponds to the zero coupling contribution, while the last one appears due to the one-loop correction by $\langle V\rangle_0$.

The one-loop partition function \eqref{z-phi} has several remarkable properties. First of all the perturbation theory in $\beta\lambda$ is well defined. Due to exponential factors in front of interaction term there are no ``ultraviolet'' divergences due to large $\omega$ contribution. Another property is that the $N$ enters in the model as a coupling for cubic interaction and as a cut-off parameter. We may conjecture that the cubic interaction term is responsible for the non planar interactions, while the planar interactions are entirely encoded into the $\varphi$ independent term in the last line of \eqref{z-phi}. Finally, the most intriguing property is ---

\section{The Hagedorn transition}

Let us evaluate the partition function \eqref{z-phi} in the leading order in $\lambda$ and $1/N$. Since we drop the cubic interaction this term becomes essentially a Gaussian integral.
The result of integration reads,
\begin{equation}\label{z-fertig}
  Z_{(1)}(\mu,x)=C\prod_{\omega=1}^{\infty}
  \frac{1}{1-\e^{-\mu_{+}\omega}-\e^{-\mu_{-}\omega}}\left(1-
  \frac{\beta\lambda}{8\pi^2}
  \frac{1}{1-\e^{-\mu_{+}}-\e^{-\mu_{-}}}+\dots
  \right).
\end{equation}

Up to notation difference and apart from U(1) contribution which we did not subtract in the partition function \eqref{z-fertig} it is precisely the same which can be obtained by counting the gauge invariant states using the P\'olya Enumeration Theorem \cite{Spradlin:2004pp}.\footnote{The identification is $\mu_{+}=\mu_{-}=\beta$.}

At zero 't Hooft coupling, $\lambda=0$, one can read from \eqref{z-fertig}, that when  the chemical potentials $\mu_{\pm}$ are approaching the critical line,
\begin{equation}\label{crit-line}
  0=1-\e^{-\mu_{+}}-\e^{-\mu_{-}},
\end{equation}
the partition function becomes singular. This singularity is due to partition function divergence which results from the dramatic increase of the density of states at small values of chemical potentials. This is the celebrated Hagedorn transition at zero coupling (see \cite{Witten:1998zw,Sundborg:1999ue,Sundborg:2000wp,%
Aharony:2003sx,Bianchi:2003wx}).

Inclusion of the one loop correction modifies the criticality condition \eqref{crit-line} to
\begin{equation}\label{crit-line-l}
  0=1-\e^{-\mu_{+}}-\e^{-\mu_{-}}+\frac{\lambda\beta}{8\pi^2}.
\end{equation}
This modification is also compatible with the results of \cite{Spradlin:2004pp}.

Now let us evaluate the effects of $N$ being finite. As we mentioned above, $N$ plays the double role: that of inverse coupling for the non-planar contribution, and one of the cut-off. An effect of finiteness of $N$ is the modification of the criticality condition \eqref{crit-line} or \eqref{crit-line-l} due to the gaussian measure for random walks and due to the cubic interaction. For large $N$, however, these effects are rather weak as soon as we did not reach the critical point.
At the critical point we have a zero mode in the Gaussian integral, but since the value of $\varphi_\omega$ is restricted to the circle: $|\varphi_\omega|\leq N$, the integral over the zero mode remains convergent and is given by the area encompassed by the circle: $\pi N^2$. This will produce a contribution to the free energy which is scaling like log of $N$.

\section{Discussions}

We have shown, that employing random walks to parameterize the partition function for the anomalous dimension matrix model in SU(2) sector greatly simplifies the analysis e.g. allowing to interpret the Hagedorn phase transition as a zero mode in the gaussian action. The transition line separates the string/chain phase of the model which is dominated by one-dimensional polymer-like configurations from the string bit/spin bit phase where there is no such a structure and the model looks like a ``soup'' of interacting spin states. In the chain phase the system has essentially no $N$ dependence, while in the other phase $N^2$ gives the effective number of particles. The physical meaning of the Hagedorn transition can be shortly described as melting of spin chain states. Let us note that at low temperatures there is another phase where the system is effectively described as one with $N$ particles (see \cite{Sochichiu:2006uz}). This phase can be conventionally called ``Higgs phase'' since it is also associated with spontaneous breaking of the gauge symmetry.

As the non-planar interactions seem to be the driving force of the transition one may ask a question: why it is possible at $N\to\infty$ when such interactions are switched off? First of all the effective rate of planar interactions was shown to be $L^2/N$, so if $N$ is extremely large but not exactly infinite\footnote{This happens if we treat $N$ as a cut off parameter.},
approaching the Hagedorn point is characterized by domination of configurations with large $L$. Then, for any large value of $N$, non-planar interactions become strong if we approach the Hagedorn point sufficiently close. On the other hand the spin chain phase does not depend on $N$ and in this case one may send $N$ to infinity from the very beginning. In this case when approaching the Hagedorn transition point the system looks overheated where configurations with very long chains tend to dominate over ones with short chains. Because of this reason the thermodynamical description of such a system fails at the Hagedorn point.
Therefore, the non-planar interactions although almost not present in the chain phase play the role of a trigger to a new thermal distribution.

\section*{Acknowledgements}
This work is made in the framework of RTN project ``Constituents, Fundamental Forces and Symmetries of the Universe''. I thank the organizers of the RTN meeting in Naples for providing an environment for lucrative communication among participants. I would like to acknowledge discussions with Pedro Silva, Edward Witten and Konstantin Zarembo.

I thank Max-Planck-Institut f\" ur Physik in Munich and in particular Dieter Luest and Fr.~Rosita Jurgeleit for warm hospitality and help. 

\end{document}